\documentclass[useAMS,usenatbib]{mn2e}
\usepackage{graphicx}
\usepackage{url}

\newcommand{\msun}{\mathrm{M}_\odot}

\title[E\"otv\"os Experiments]{E\"otv\"os Experiments with 
  Supermassive Black Holes  }
\author[Asvathaman, Heyl \& Hui]{Asha Asvathaman$^{1}$, Jeremy S. Heyl$^{1}$\thanks{Email:
    heyl@phas.ubc.ca; Canada Research Chair}, Lam Hui$^{2}$ \\
$^{1}$Department of Physics and Astronomy, University of British
  Columbia, 6224 Agricultural Road, Vancouver, BC V6T 1Z1, Canada\\
  $^{2}$Physics Department, Institute for Strings, Cosmology, and Astroparticle Physics, Columbia University, New York, NY 10027, USA}

\begin{document}
\date{Accepted ---. Received ---; in original form ---}

\pagerange{\pageref{firstpage}--\pageref{lastpage}} \pubyear{2015}

\maketitle

\label{firstpage}

\begin{abstract}
  By examining the locations of central black holes in two elliptical
  galaxies, M\,32 and M\,87, we derive constraints on the violation of
  the strong equivalence principle for purely gravitational objects,
  {\em i.e. black holes}, of less than about two-thirds, $\eta_N<0.68$
  from the gravitational interaction of M\,87 with its neighbours in
  the Virgo cluster.  Although M\,32 appears to be a good candidate
  for this technique, the high concentration of stars near its centre
  substantially weakens the constraints.  On the other hand, if a
  central black hole is found in NGC~205 or one of the other satellite
  ellipticals of M\,31, substantially better constraints could be
  obtained. In all cases the constraints could improve dramatically
  with better astrometry.
\end{abstract}

\section{Introduction}

The strong equivalence principle (SEP) states that any object
regardless of its composition will travel through a gravitational
field in the same way.  This includes objects with varying
contributions of gravitational energy.  In particular a black hole
whose mass is entirely gravitational should travel in the same manner
through the gravitational field as a
star. \citet{1968PhRv..169.1014N,1968PhRv..169.1017N} argued that
metric theories of gravity other than general relativity may exhibit
violations of the SEP --- in particular,
objects 
which
have a large contribution of gravitational energy to
their makeup may travel differently through a gravitational field than
other objects.

Typically the strong equivalence principle is probed by looking for
the Nordvedt effect in the Earth-Moon system or binary stars
consisting of a neutron star and a white dwarf \citep{Stairs:2005}.
The \citet{1968PhRv..170.1186N} effect results in the polarization of
an orbit in the direction of the gravitational acceleration of a large
body: the Sun in the case of the Earth-Moon system
\citep{1976PhRvL..36..551W,1976PhRvL..36..555S}, and the Galaxy in the
case binary pulsars.  The recently discovered millisecond pulsar in a
triple system, PSR J0337+1715, will provide further interesting
constraints \citep{Ransom:2014aa}.

Here we perform the test proposed by \cite{2012PhRvL.109e1304H}.
In particular we will examine the polarization of the orbits of
supermassive black holes through the central regions of elliptical
galaxies.  We will focus on elliptical galaxies because the presence
of central black holes is common in sufficiently large elliptical
galaxies and the location of the bottom of the gravitational potential
is straightforward to constrain by determining the centroid of the
distribution of stellar light.  Interactions between galaxies are
ubiquitous, so example systems are straightforward to find.  In
particular we will examine the small elliptical galaxies that orbit
the largest neighbour of the Milky Way galaxy, the Andromeda galaxy or
M\,31, and the massive elliptical galaxy, M\,87, in the Virgo cluster.
We will derive an upper limit on SEP violation -- how this limit
should be interpreted in the context of general gravitational theories
will be discussed in \S \ref{conclusions}.

\section{Calibration}

Let us suppose that we are studying a small stellar system in the
gravitational field of a large one.  The stars in the small system
experience an acceleration toward the centre of the larger one
\begin{equation}
  a_\star = \frac{GM}{d^2}
  \label{eq:1}
\end{equation}
where $d$ is the distance between the small system and the larger one
and $M$ is the mass of the larger system.  These accelerations are
typically ten to one hundred times larger than those exerted by
large-scale structure $a \sim 600~\mathrm{km~s}^{-1} H_0 \sim 10^{-10}
\mathrm{cm~s}^{-2}$.

Furthermore, let us assume that a massive black hole within the
smaller system experiences a different acceleration
\begin{equation}
  a_\bullet = \left ( 1 - \Delta \right ) a_\star
  \label{eq:2}
\end{equation}
where $\Delta$ quantifies the extent of violation of the
SEP. \citet{1982RPPh...45..631N} presents an equivalent definition
where the ratio of the inertia masses and passive gravitational mass
of an object could depend on its constitution and in particular the
contribution of gravitational energy,
\begin{equation}
  \frac{m_\mathrm{Gp}}{m_\mathrm{I}} = 1- \eta_N \frac{U_\mathrm{G}}{mc^2}.
  \label{eq:3}
\end{equation}
In the context of the parametrized post-Newtonian treatment of
deviations from general relativity \citep{Will:lrr},
\begin{equation}
  \eta_{N} =4\beta - \gamma - 3 - \frac{10}{3} \xi - \alpha_1 +
  \frac{2}{3} \alpha_2 - \frac{2}{3} \zeta_1 - \frac{1}{3} \zeta_2.
  \label{eq:4}
\end{equation}
For the case of a black hole, $U_\mathrm{G} = mc^2$ so
$\Delta=\eta_N$. 

The massive black hole also experiences an acceleration due to the
stellar system in which it resides.  Furthermore, dynamical friction
with the less massive stars nearby damps any motion of the black hole
relative to its equilibrium position on a short time scale, so absent
SEP violation or rare astrophysical effects such as a recent merger or
asymmetric jets the black hole should lie at the bottom of the
gravitational potential well.  The effects of individual stellar
encounters, Brownian motion, are much smaller than what we discuss
here \citep{2011ApJ...735...57B}.

Near the minimum of the stellar potential we can approximate it as a
harmonic oscillator so
\begin{equation}
  \left (1 - \Delta \right) \omega^2 x
   = a_\bullet - a_\star = -\Delta \frac{GM}{d^2}
  \label{eq:5}
\end{equation}
and $x$ is the displacement of the black hole from the centre of the
small stellar system in the direction of the larger system. If
$\Delta>1$ the black hole would not lie near the centre of the galaxy,
so this region is excluded.

The value of $\omega$ depends on the structure of the small stellar
system and depends on the mean density within the region surrounding
the black hole.  For a \citet{1990ApJ...356..359H} model for the
density distribution that well describes the stellar distribution of
elliptical galaxies
\begin{equation}
  \rho = \frac{m}{2\pi} \frac{a}{r} \frac{1}{(r+a)^3}
  \label{eq:6}
\end{equation}
where $R_\mathrm{eff}$, the half-light radius in projection is
approximately $1.8153 a$.  We take the black hole mass to be
$m_\bullet$ to be much less than that of the small system.  However,
within a small region near the black hole, the mass of the black hole
dominates that of the stars.  We will take the mass of the stars
within this sphere of influence to be $\alpha m_\bullet$ where
$\alpha\approx 1$.  Now if the black hole is displaced from the centre
by violation of the equivalence principle, Eq.~\ref{eq:5}, then we
assume that these stars will be displaced with the black hole and
remained centred on the black hole with a Hernquist distribution of
density, Eq.~\ref{eq:6}.  On the other hand, we will assume that the
rest of the galaxy will remained centred on the undisplaced position
of the black hole.  Of course, there will be a smooth transition
between these two regimes, but this approximation contains the
essential point that the cusp of stars at the centre of the galaxy
will remain centred on the black hole.  By numerically integrating
over the gravitational potential energy of this configuration
we find 
\begin{equation}
  \omega^2 \approx 0.30 \left ( \frac{m}{\alpha m_\bullet} \right )^{1/2} \frac{Gm}{a^3}
  \label{eq:7}
\end{equation}
where the coefficient comes from the numerical integration and the scaling comes
from considering the structure of the Hernquist model at small radii.

If one directly used Eq.~(\ref{eq:6}) to calculate the restoring force
toward the centre, one would find that it is constant with
displacement from the centre.  This would assume that the density is
actually singular at the centre but also would neglect the effect of
the black hole on the neighbouring stars which would be dragged with
it.  We will compare the estimate including the effect of neighbouring
stars (Eq.~\ref{eq:7}) with the detailed measurements of the potential
near the core of the elliptical galaxy, M\,87, as inferred from
stellar velocities and find good agreement.

We will take $\alpha=1$ and $m_\bullet=0.005m$ where 
$m$ is the mass of a bulge and
halo approximately modelled by a Hernquist model \citep[as
  in][]{1997ApJ...487..153V}, we find that
\begin{equation}
  \frac{x}{d} = 0.24 
\frac{\Delta}{1-\Delta} \frac{M}{d^3} \frac{a^3}{m},
  \label{eq:8}
\end{equation}
so the relative displacement only depends on the relative masses and
sizes of the various bodies and the angular displacement does not
depend on the distance to the systems from Earth.  We can make this
explicit by calculating the apparent displacement across the sky as
\begin{equation}
  \Delta \theta = 0.24 \frac{\Delta}{1-\Delta} \frac{M}{\delta^2} \frac{\alpha^3}{m} \cos^3 i
\end{equation}
where $i$ is the inclination of the line connecting the two bodies
with respect to the plane of the sky, $\delta$ is the angular distance
between the bodies and $\alpha$ is the angular size of the Hernquist
radius $a$ of the host galaxy.  If smaller galaxies follow
an isothermal distribution about the perturbing galaxy as
\citet{2006AJ....131.1405K} found for the satellites of M\,31, the
mean value of the geometric term $|\cos^3i|$ is $4/(3\pi) \approx
0.42$ and the median value is $\sqrt{2}/4\approx 0.35$.  Ninety
percent of the time the term lies between $5\times 10^{-4}$ and 0.99,
so geometric information is especially useful to give firm
constraints.  Of course the halo of the galaxy group must end
somewhere so the lower limit on $\cos^3 i$ found here is somewhat
unrealistic.  On the other hand if the density of satellites is
proportional to $r^{-3}$ like the outer regions of the
Navarro-Frenk-White profile (\citeyear{1996ApJ...462..563N}), the mean
value of the geometric term is $3\pi/16 \approx 0.59$ and the median
value is $3\sqrt{3}/8 \approx 0.65$.  Ninety percent of the time the
term lies between 0.03 and 0.996.

\section{The Andromeda Galaxy System}

\begin{figure}
  \includegraphics[width=\columnwidth,clip,trim=2.5in 7in 2.5in 1.7in]{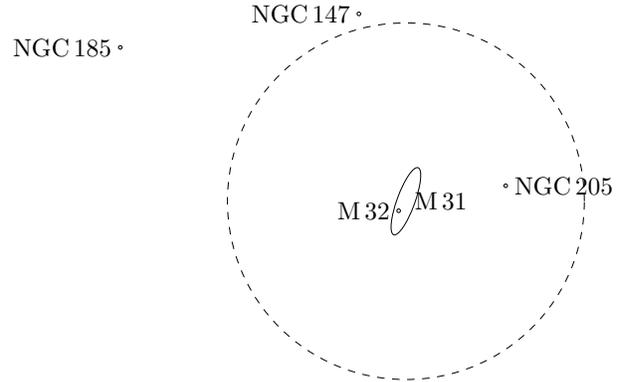}
  \caption{The geometry of the largest elliptical satellite galaxies
    of the Andromeda Galaxy. The Sun is to 773~kpc to the left of
    M\,32, and the radius of the dashed circle is 100~kpc.}
  \label{fig:M31-system}
\end{figure}

The nearest neighbour of the Milky Way has four small elliptical
galaxies orbiting it: M\,32, NGC\,205, NGC\,147 and NGC\,185.
Fig.~\ref{fig:M31-system} depicts the location of the various galaxies
relative to each other from the distances compiled by
\citet{2006AJ....131.1405K}.  Of course, the distances to the various
galaxies are more uncertain than their positions on the sky.

\begin{figure}
  \includegraphics[width=\columnwidth]{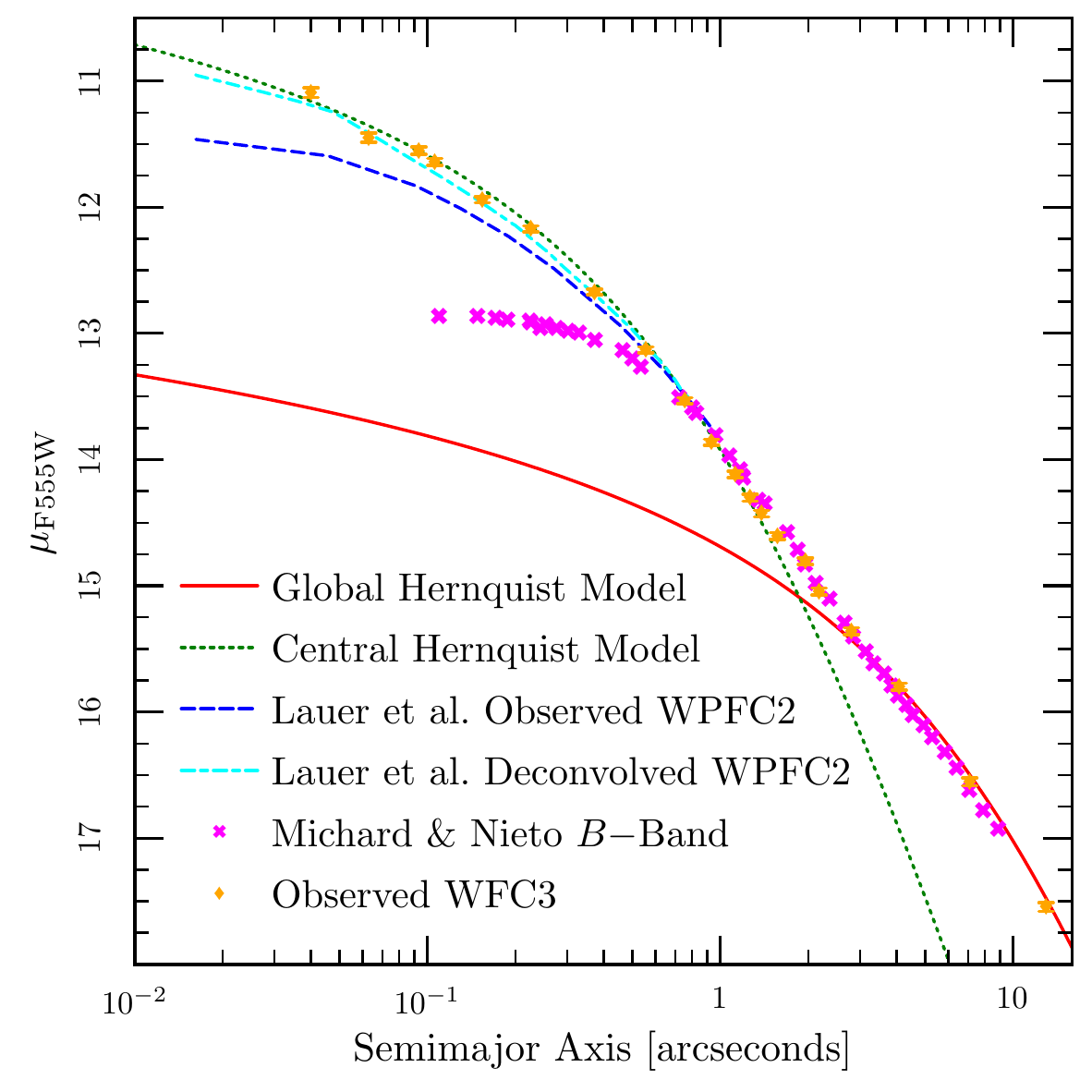}
  \caption{The surface brightness in the core of M\,32 from
    \citet{1998AJ....116.2263L}, \citet{1991A&A...243L..17M}
    and the photometry and models used in this paper.  We assumed
    $B-F555W=1$ to place the \citet{1991A&A...243L..17M} data
    on this plot.}
  \label{fig:M32-iso}
\end{figure}

A supermassive black hole (SMBH) has already been discovered in the
core of the compact elliptical galaxy M\,32
\citep{1997Natur.385..610V} and found to be within 0.5 arcseconds of
the nucleus of the galaxy \citep{2015arXiv150203231Y}.  M\,32 is
located about 0.4$^\circ$ from the centre of M\,31 in the plane of the
sky.  We can estimate the displacement of the black hole from the
centre of M\,32 as a function of $\Delta$ in the best-case scenario of
the minimal separation of M\,32 from M\,31 of about 5~kpc.  In this
case M\,32 would be well within the halo of M\,31, and we could use
the circular velocity of M31 of 200~km/s at this distance to estimate
the acceleration of M\,32 toward the centre of M31, $a_\star$ in
Eq.~(\ref{eq:1}),
\begin{equation}
  a_\star = \frac{v^2}{d} = 3.6\times 10^{-8} \mathrm{cm~s}^{-2}
  \label{eq:9}.
\end{equation}
For the galaxy M\,32 as a first approximation we will use a Hernquist
model that fits to the global structure of the galaxy.  This model
fits the observed half-light radius of the galaxy (we will revisit
this assumption later in light of data that probe the centralmost
regions, see Fig.~\ref{fig:M32-iso}), so we take $m=8 \times 10^8 \msun$
and $R_\mathrm{eff} = 100$~pc so $a=55$~pc and
\begin{equation}
  x = 0.091 \frac{\Delta}{1-\Delta} \mathrm{pc}
  \label{eq:10}
\end{equation}
or
\begin{equation}
  \Delta \theta = \frac{x}{d_\mathrm{M32-Sun}} = 0.0288 \frac{\Delta}{1-\Delta} \mathrm{arcseconds},
  \label{eq:11}
\end{equation}
so if M\,32 lies indeed at the close to minimum distance to M\,31, the
existing measurement of the position of the black hole yields a
constraint of $\Delta < 0.95$.

We can improve upon the astrometric constraint of
\citet{2015arXiv150203231Y} if we assume that the position of the
black hole is coincident with the cusp in the surface brightness of
the galaxy.  This is a reasonable assumption because the black hole
would entrain the neighbouring stars, so the stellar nucleus should be
centred on the black hole. To measure the possible deviation of the
position of the black hole on the sky and the centre of the potential
well of the galaxy we use images from the Wide-Field Camera 3 (WFC3)
on the Hubble Space Telescope (HST) in the filter, F555W.  This
instrument features 0.04~arcsecond pixels and a $162 \times
162$-arcsecond field of view.  Both the Space Telescope Imaging
Spectrograph (STIS) and the planetary camera on Wide-Field Planetary
Camera 2 (WF/PC2) feature a finer pixel scale but at the expense of
field of view.  We used images from GO-11714 (PI: Bond) with total
exposure time of 120~seconds.  To localise the stellar core which lies
within the black hole sphere of influence of about 0.3~arcseconds
\citep{2001ApJ...550..668J}, we used circular and elliptical apertures
of 0.16 arcseconds and 0.2 arcseconds.  To localise the centroid of
the outer regions we used two elliptical annuli: one of inner axes 2.8
and 4.3 arcseconds and outer axes of 5.7 and 8.5 arcseconds and one
half that size, well outside the black hole's sphere of influence.
The centres of the annuli are initially roughly placed within about an
arcsecond of the core and updated to lie on the calculated
centroid. This is iterated thirty times.  The centroids of the various
annuli to probe the galactic potential coincide to within
0.01~arcseconds as do the ellipses and circles centred on the stellar
cusp, yielding an estimate of the precision of 0.01~arcseconds.
Furthermore, the centroid of the cusp and the outer regions also agree
to within 0.01~arcseconds.  If we combine this constraint with the
distance estimate of \citet{2006AJ....131.1405K}, this yields a
constraint $-0.53<\Delta<0.26$.  In fact the maximal distance between
M\,31 and M\,32 consistent with the uncertainties given by
\cite{2006AJ....131.1405K} is 45~kpc \citep[see also][for an
  alternative interpretation]{2014ApJ...788L..38D}.  In the case the
constraints on $\Delta$ would be much weaker with $\Delta<0.99$.

Before we focus on the smaller elliptical galaxies of the M\,31
system, we will examine the central region of M\,32 in further detail.
\citet{1998AJ....116.2263L} deconvolved images with HST and WFPC-1 and
WFPC-2 and found a very high central concentration of stars.
Fig.~\ref{fig:M32-iso} depicts the surface brightness in the core of
M\,32 from ground-based data \citep{1991A&A...243L..17M} and HST WFPC2
images \citep{1998AJ....116.2263L} as well as the WFC3 images and the
models that we are using in this paper.  We can see that the global
Hernquist model that reflects the half-light radius of the galaxy of
about 27~arcseconds dramatically underpredicts the light in the core
of the galaxy.  On the other if we focus on the inner two arcseconds
of the galaxy we can characterise the distribution of light with a
much more concentrated Hernquist model with $a=1.3$~arcseconds or
5.5~pc.  The total mass in this central Hernquist model is about 0.17
of the entire galaxy or $1.4\times 10^8\msun$.  This more concentrated
model underpredicts the light outside of two arcseconds, but it does
well in the region of the central black hole whose mass we have
assumed to be about $4\times 10^6\msun$, still significantly smaller
than the mass of the central Hernquist model, so we can apply the
preceding analysis here but with the new parameters for the Hernquist
model.

If we examine Eq.~\ref{eq:7}, we see that the spring constant of the
Hernquist model that best reflects the central region of M\,32 is
about 70 times larger than that of the model of the entire galaxy and
the expected displacement is 70 times smaller,
\begin{equation}
  \Delta \theta = 0.4 \frac{\Delta}{1-\Delta} \textrm{milliarcseconds}.
\end{equation}
The current astrometry yields $\Delta \theta<10$~milliarcseconds,
yielding a weaker constraint of $\Delta<0.96$.  

The three smaller elliptical galaxies, NGC\,205, NGC\,147 and NGC\,185,
have smaller mass densities and may have more favourable geometry.  However,
SMBHs have not yet been identified in these galaxies.  Each of these
galaxies has a mass of about $2 \times 10^8 \msun$ and a
$R_\mathrm{eff} \approx 300$~pc, so $a \approx 165$~pc.  The two
galaxies, NGC\,147 and NGC\,185 are each about 100~kpc from M31, so the
enclosed halo mass is about $10^{12} \msun$.  This yields a
typical value of the displacement of
\begin{equation}
  x = 0.55 \frac{\Delta}{1-\Delta} \mathrm{pc} ~\mathrm{and}~
  \Delta \theta = 0.18 \frac{\Delta}{1-\Delta} \mathrm{arcseconds}.
  \label{eq:12}.
\end{equation}
Although NGC\,205 may be somewhat closer to M~31 at 50~kpc than the
other galaxies, the geometry is somewhat poorer according to the
distance given by \citet{2006AJ....131.1405K} perhaps reducing the
angular displacement by a factor of two counteracting the increased
acceleration.  This yields a similar observed displacement.  If a SMBH
is discovered near the centre of these galaxies and can be localised
within 0.5~arcseconds, this would yield a constraint of $\Delta<0.74$.

On the other hand, NGC\,205 may lie at the same distance from
us as M\,31. This is consistent with the errorbars given by
\citet{2006AJ....131.1405K}.  Furthermore, \citet{2006AJ....131..332G}
argue that NGC\,205 is interacting tidally with M\,31, favouring this
interpretation.  At this minimum possible distance to M\,31 of about
7~kpc, and we obtain a much larger linear and angular displacement
\begin{equation}
  x = 10. \frac{\Delta}{1-\Delta} \mathrm{pc} ~\mathrm{and}~
  \Delta \theta = 3.1 \frac{\Delta}{1-\Delta} \mathrm{arcseconds}.
\end{equation}
In this most favourable case the constraint would be $-0.19 < \Delta < 0.14$
with 0.5-arcsecond astromety and $|\Delta|<3\times 10^{-2}$ with
0.01-arcsecond astrometry.

\section{Virgo Elliptical Galaxies}

One of the largest elliptical galaxies in the Virgo Cluster, M\,87,
also contains a supermassive black hole and lies at the centre of the
Virgo A subcluster which contains the bulk of the mass of the Virgo
Cluster, about $10^{14} \msun$.  The Virgo Cluster is still in the
process of forming and contains at least two additional subclumps of
masses of about $10^{13} \msun$.  The subclump closest to M\,87 is
centred on M\,84 and M\,86 about $1.3^\circ$ away.  M\,89 is also a
similar angular distance away, so this analysis applies to it as well,
but it is likely to be less massive.

\begin{figure}
  \includegraphics[width=\columnwidth]{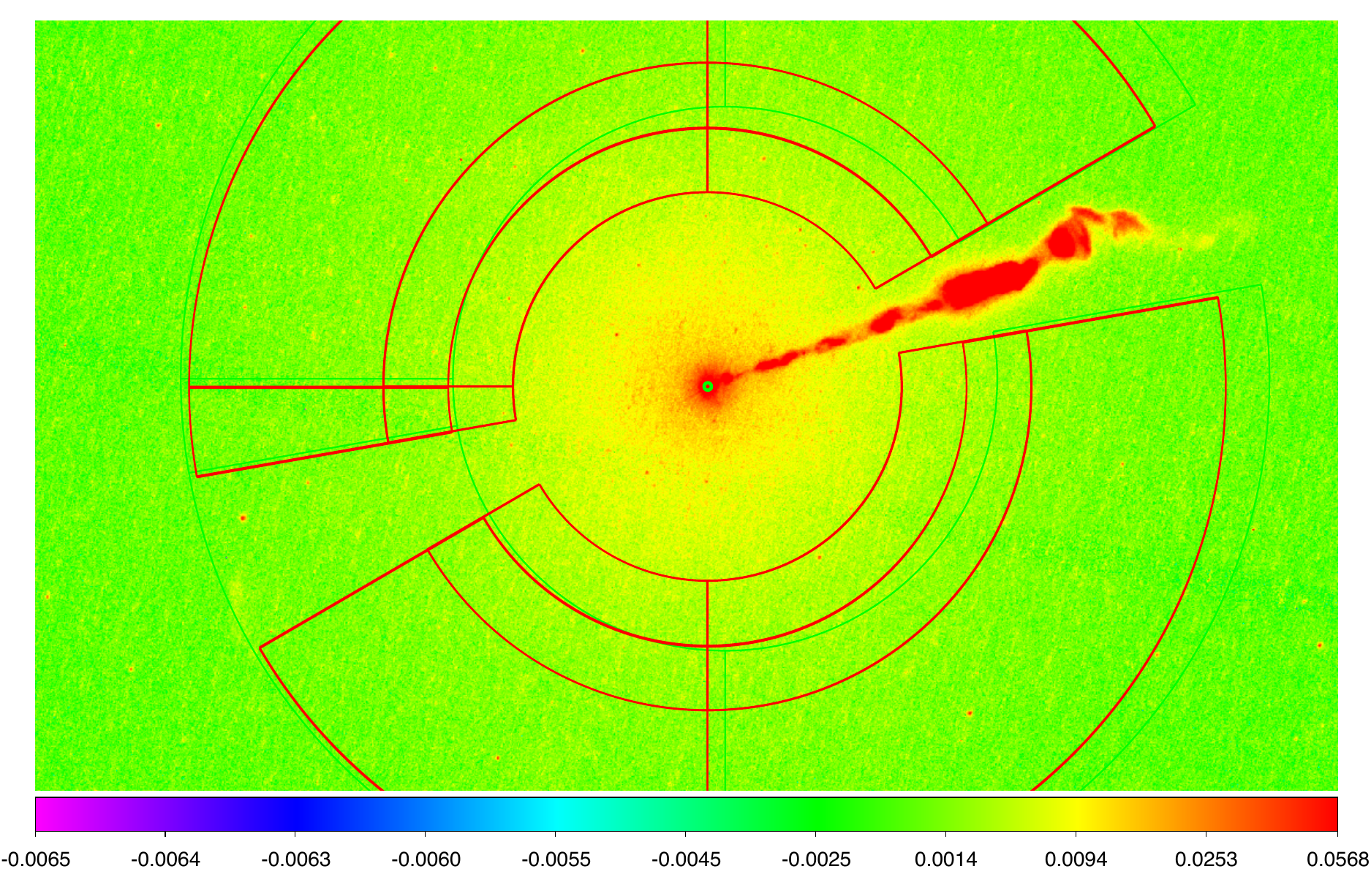}
  \caption{Central region of M\,87 focusing on the AGN.  Several
    apertures to determine the centroid of the light are depicted.
    The two middle annuli lie from 9 to 15 arcseconds and 12 to 24
    arcseconds from the centre.  The inner apertures are circular and
    0.15 and 0.2 arcseconds in radius. The outermost aperture between
    24 and 36~arcseconds is not depicted.  The galaxy M~89 is east
    (left) of M~87 and M~84 is to the West (right).}
  \label{fig:M87}
\end{figure}

\begin{figure}
  \includegraphics[width=\columnwidth]{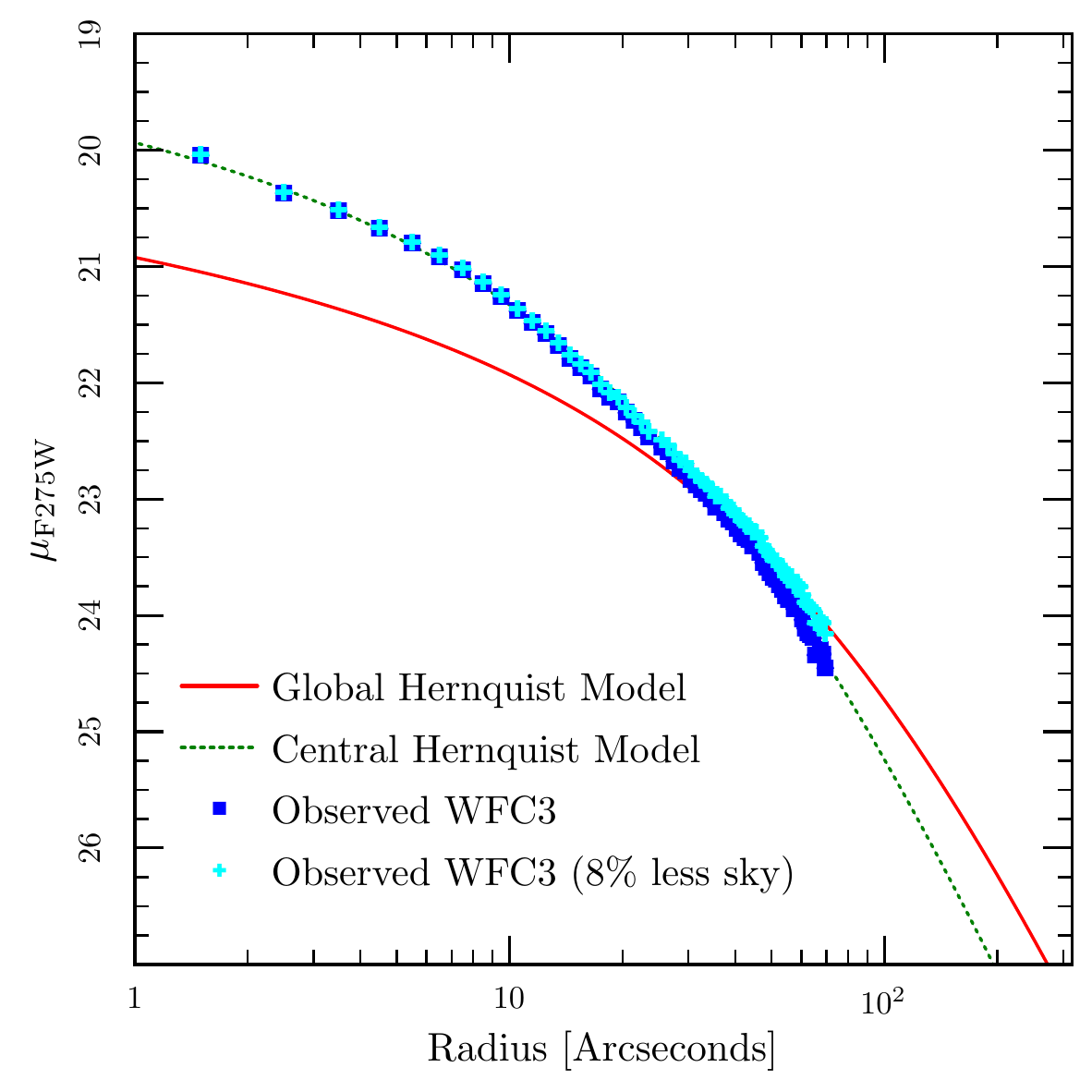}
  \caption{The surface brightness in the core of M\,87 as measured
    from the image depicted in Fig.~\ref{fig:M87} along with the
    Hernquist model based on the measured $R_\mathrm{eff}$ of
    $164^{\prime\prime}$ and one based on the inner F275W photometry
    with $R_\mathrm{eff}$ of $74^{\prime\prime}$.}
  \label{fig:M87-iso}
\end{figure}

For the galaxy M\,87 we take $R_\mathrm{eff}=164^{\prime\prime}$
\citep{2006ApJS..164..334F} so $a=90^{\prime\prime}$.  We will take
the mass of M\,87 to be about $4\times 10^{12} \msun$
\citep{2006ApJ...643..210W} and take the mass of the subclump
containing M\,84 and M\,86 to be $10^{13} \msun$ at a distance of
$1.3^\circ$. This yields a displacement of
\begin{equation}
  \Delta \theta = 0.02 \cos^3 i \frac{\Delta}{1-\Delta} 
  \mathrm{arcseconds}.
  \label{eq:13}
\end{equation}
We can examine the stellar distribution of M\,87 in further detail.
We use images from the Wide-Field Camera 3 (WFC3) on the Hubble Space
Telescope (HST) in the ultraviolet filter, F275W.  We used images from
GO-12989 (PI: Renzini) with total exposure time of 5599~seconds as
depicted in Fig.~\ref{fig:M87}.  M\,87 is an E0 galaxy so its
isophotes are nearly circular \citep{2006ApJS..164..334F}, so to
measure the surface brightness we average over circular annuli of
width of one arcsecond that exclude the jet and the direction opposite
the jet as depicted by the regions in Fig.~\ref{fig:M87} (but with
larger widths).  Fig.~\ref{fig:M87-iso} depicts the resulting surface
brightness profile that extends to 70 arcseconds, the edge of the
image.  This is well within the half-light radius of the galaxy, so we
cannot probe the distant light well, and the surface brightness
measured at large radii depends on the assumed sky brightness which we
also fit in parallel with the Hernquist model.  We again see that the
global Hernquist model based on the half-light radius does not fit the
central region well.  \citet{2006ApJS..164..334F} find that the
central region of M\,87 is better fit by a ``core-S\'ersic'' model
\citep{1968adga.book.....S,2004AJ....127.1917T} that has a much less
cuspy central surface brightness distribution than a Hernquist model.
The break radius between the core and the outer S\'ersic model is
about 7~arcseconds.  Furthermore, in the core of M\,87 we have
detailed kinematic data, so we can estimate the properties of the
gravitational potential induced by the stars.

\citet{2013ApJ...770...86W} determine the mass of the central black
hole of M\,87 to be about $3.5 \times 10^{9} \msun$ (approximately
$10^{-3}$ of the mass of the galaxy) and to dominate the mass 
within about 5.61~arcseconds of the centre.  They also model the
contribution of the stellar mass to the circular velocity in this
inner region, yielding a value of about 160~km/s at $4^{\prime\prime}$
from the black hole consistent with the photometry of
\citet{2006ApJS..164..334F} and a mass-to-light ratio of 4. Assuming a
smaller mass-to-light ratio would yield smaller 
constraints 
on $\Delta$, roughly the expected displacement for a given value of
$\Delta$ is inversely proportional to the square of the mass-to-light
ratio.  As the stellar mass-to-light ratio decreases, the effect of
the stellar potential diminishes, increasing the displacement, and the
sphere of influence of the black hole increases, so we can probe the
effect at larger radii from the black hole, further increasing the
expected displacement.

Within 10~arcseconds the black hole contributes about thirty percent
of the mass, so we will use the circular velocity due to the stellar
contribution at this radius of about 210~km/s to estimate the
displacement of the black hole due to SEP violation.  Using a distance
to M\,87 of 16.4~Mpc \citep{2010A&A...524A..71B} and the same mass and
angular distance to the subclump centred on M\,84 as earlier yields
\begin{equation}
  x = 1.1 \frac{\Delta}{1-\Delta} \cos^2 i ~\mathrm{pc}
  \label{eq:14}
\end{equation}
and
\begin{equation}
    \Delta \theta = 0.014 \frac{\Delta}{1-\Delta} \cos^3 i~
    \mathrm{arcseconds}
    \label{eq:15}
\end{equation}
where we have extrapolated the theoretical stellar mass profile of
\citet{2013ApJ...770...86W} out to 10~arcseconds, slightly beyond the
break radius of 7.15~arcseconds determined by
\cite{2006ApJS..164..334F} where the surface brightness profile
steepens.


The position of the black hole can in principle be determined to
microarcsecond precision with microwave interferometry
\citep{2011ApJ...735...57B}.  On the other hand, the centre of the
potential well is best defined in the optical using the isophotes of
the galaxy.  Fortunately, the supermassive black hole is also apparent
in the optical.  The innermost isophotes will be centred on the black
hole because it will dominate the potential well, so the centres of
more distant isophotes provide an estimate of the potential well that
constrains the black hole.

\citet{2010ApJ...717L...6B} argued from isophotal analysis of
observations with the Advanced Camera for Survey on HST that the AGN
may be displaced from the centre of the galaxy by about 7~pc in a
direction opposite to the observed jet (about 0.1~arcseconds).
However, subsequently 
neither
\citet{2011ApJ...729..119G} nor \citet{2013ApJ...770...86W} found
evidence for displacement greater than about one parsec; the position
determined by \citet{2011ApJ...729..119G} was consistent with that of
\citet{2010ApJ...717L...6B} within the larger errorbars of
\citet{2011ApJ...729..119G}.  \citet{2010ApJ...717L...6B} outlined
several astrophysical explanations for a potential displacement such
as a SMBH binary, a recent merging of black holes or a one-sided jet.
The interaction with the neighbouring stars or even clusters of stars
is too weak to explain such a displacement.

To measure the possible deviation of the position of the black hole on
the sky and the centre of the potential well of the galaxy
we 
calculated 
the light centroid within three circular annuli of inner
and outer radii 24 and 36~arcseconds, 12 and 24~arcseconds, and 9 and
15~arcseconds.   If we take the mass of the
black hole to be about $4\times 10^9\msun$, it will dominate the mass
within about 6.6~arcseconds from the centre.  Within 12~arcseconds it
contributes less than 25\% of the mass, and within 24~arcseconds it
contributes about 8\% of the mass; therefore, these annuli lie outside
the black hole's sphere of gravitational influence.  The slices of the
annulus along the direction of the observed jet and potential
counterjet are omitted from the calculation and so exclude light from
the AGN itself.  This 
technique 
contrasts with that of
\citet{2010ApJ...717L...6B}.  They explicitly masked the noticeable jet
emission as well as globular clusters.  Here we remove a much larger
region from the analysis both in the direction of the jet and the
opposite direction.  This excludes both the observed jet and a wide
regions around the jets.  The excluded region is symmetric, so we
minimize any potential bias along the jet axis.  Furthermore, the
measurements that we use are in F275W with a higher angular resolution
camera (WFC3 vs. ACS).  In this ultraviolet filter, the emission from
globular clusters is negligible.  The centre of the annuli are
initially roughly placed within about an arcsecond of the AGN and
updated to lie on the calculated centroid. This is iterated thirty
times.

To determine the location of the black hole we performed a similar
centroiding on circular
apertures
of 0.15 and 0.2 arcseconds, also initially centred on the AGN within
0.2 arcseconds.  We have repeated this process several times using
different starting positions, resulting in slightly different
centroids all consistent within their mutual standard deviation of
0.03~arcseconds, slightly less than a pixel.  This is consistent with
the precision found by \citet{2013ApJ...770...86W} and poorer than we
found with M~32 because the surface brightness profile of M\,87 is
much shallower.  The isophotes constrain the deviation of the black
hole from the centre of the potential within 0.03~arcseconds, yielding
a constraint on the violation of the SEP for black holes of
$\Delta<0.68$, assuming the more conservative displacement estimate
Eq.~(\ref{eq:15}) and a favourable geometry of $i\approx 0^\circ$.
Using additional images, more sophisticated isophote fitting procedure
and more detailed modelling of the mass distribution of the Virgo
cluster could possibly yield smaller constraints.

\section{Conclusions}
\label{conclusions}

How does our limit on $\Delta$ fit within the context of existing
scalar-tensor theories? SEP violation occurs in these theories because
compact objects have a suppressed scalar charge, leading to a weaker
scalar coupling to the environment. For the classic Brans-Dicke
theory, solar system tests constrain the scalar-matter coupling to be
fairly weak to begin with (much weaker than tensor-matter), thus
predicting SEP violation that is much below our limit.  More recent
scalar-tensor theories open up the interesting possibility of
predicting observable SEP violation on large scales while respecting
the solar system constraints -- the latter is achieved by what is
known as screening mechanism. Screening introduces a subtlety in the
interpretation of $\Delta$, however -- it becomes scale/environment
dependent. In the case of the M32-M31 system, we estimate $\Delta \sim
10^{-3}$ in galileon theories \citep{2009PhRvD..79f4036N}.  For such
theories, $\Delta \sim 1$ can be found in situations where the
acceleration is due to large scale structure, quite a bit smaller than
the acceleration of our setup \citep{2012PhRvL.109e1304H}.  Despite
this, deriving a phenomenological limit on $\Delta$ is useful, for
several reasons.\footnote{Note also the environmental dependence of
  $\Delta$ is often such that $\Delta$ on the left hand side of Eq.~(\ref{eq:5})
  is much smaller than $\Delta$ on the right; the statements above
  refer to the latter.}  First, new theories -- massive gravity being
an example -- are still being developed. Second, equivalence principle
violation remains to be worked out even for some existing theories
\citep{2009PhRvD..80j4002H}.  Third, this type of galaxy-satellite
system constraint on $\Delta$ can be improved.

Both the calibration and observational analysis presented here can be
improved upon in several directions.  From the calibration standpoint,
we have 
considered
just the force on the black hole and the coterie
of stars in its immediate gravitational influence.  Moving the black
hole from the centre of the stellar system will also move these stars,
and we have used a simple model for their coupled motion.
Additional study is warranted where one would start the black hole
in the centre of the galaxy increase the differential acceleration gradually,
say due to the effect of the small galaxy 
approaching 
the larger one, and determine where the black hole ends up.

As we have mentioned earlier, there are more sophisticated ways to
determine the centre of the galaxy with isophotal fitting and by
considering measurements from several observations at several
wavelengths.  In 
principle 
one could find tighter constraints on the deviation of the black hole
from the centre of the potential.  In the case of M\,32 the existing
constraints are just 0.5~arcseconds because one has to link the radio
coordinate system to the visual one.  These could possibly be improved
dramatically with further observations.  The visual core of M\,32 lies
within 0.01~arcseconds of the centroid of the outer isophotes of the
galaxy, indicating that if the black hole is coincident with the core,
$\eta_N<0.96$, using a density model adapted to fit the central
regions of M\,32.  The discovery of a supermassive black hole in one
of the other elliptical satellites of M\,31 could provide stronger
constraints from the M\,31 system.  However, the main uncertainty in
the constraints comes from the unknown geometry of the system and the
central potential as we demonstrate for M\,32 in particular so further
information on the relative distances and the central kinematics of
the various galaxies would yield better constraints on the results.
Observations of the black hole in M\,87, the stellar kinematics near
the black hole and an assumption of a favourable geometry for the
neighbouring galaxies within the Virgo cluster yields a constraint of
$\eta_N<0.68$, competitive with the results from pulsar timing
\citep{Stairs:2005}.  Perhaps other interacting and nearly interacting
galaxies in the local Universe could also provide more information and
would statistically limit the violation of SEP in the motion of black
holes.

The constraint on SEP from the motion of black holes probes the SEP 
in an essentially different limit 
from that of
lunar ranging; however, the technique outlined here could become
competitive with the lunar ranging results which now constrain
$|\eta_N|<1.3 \times 10^{-3}$ \citep{Baess:1999,2001CQGra..18.2397A}.
Improved astrometric measurements of the centre of the potential of
M\,87 and the location of the black hole could be achieved with
existing visual and UV data and the position of the black hole could
be
constrained 
further with radio 
interferometry 
provided
the optical and radio astrometry could be tied together with
sufficient precision.  Relative astrometry with a precision $\sim
10^{-4}$ arcseconds can be achieved with Hubble photometry
\citep[e.g.][]{Heyl116397dyn}, a factor of one hundred better than
presented here, so constraints on $|\eta_N|$ of $10^{-3}$ could be
possible with supermassive black holes.  If microarcsecond astrometry
can be brought to bear \citep[e.g.][]{2011ApJ...735...57B} on M\,87,
then 
constraints
on $\eta_N$ on the order of $10^{-4}$ might be possible.

{\noindent \bf Acknowledgements}

This research is based on NASA/ESA Hubble Space Telescope observations
obtained at the Space Telescope Science Institute, which is operated
by the Association of Universities for Research in Astronomy
Inc. under NASA contract NAS5-26555 and accessed through the Hubble
Legacy Archive which is a collaboration between the Space Telescope
Science Institute (STScI/NASA), the Space Telescope European
Coordinating Facility (ST-ECF/ESA) and the Canadian Astronomy Data
Centre (CADC/NRC/CSA). These observations are associated with proposal
GO-11714 (PI: Bond) and GO-12989 (PI: Renzini).  This work was
supported by the Natural Sciences and Engineering Research Council of
Canada, the Canadian Foundation for Innovation, the British Columbia
Knowledge Development Fund, the Bertha and Louis Weinstein Research
Fund at the University of British Columbia, the Department of Energy
and NASA. LH thanks Henry Tye and the HKUST Institute for Advanced
Study for hospitality.  The authors would also like to thank the
anonymous referee for detailed comments that improved the paper
substantially.

\bibliography{weighbh}
\bibliographystyle{apj}

\label{lastpage}
\end{document}